\documentclass[10pt,twocolumn,journal]{IEEEtran}
\normalsize
\usepackage{times}
\usepackage{amsbsy}
\usepackage{amssymb}
\usepackage{fontenc}
\usepackage[usenames,dvipsnames]{color}
\usepackage{latexsym}
\usepackage{amsmath}
\usepackage[ansinew]{inputenc}
\usepackage[dvips ]{graphicx}
\input epsf
\usepackage{caption}
\usepackage[caption=false,font=footnotesize]{subfig}
\begin{document}
\title{\LARGE Coordinate Tomlinson-Harashima Precoding Design for Overloaded Multi-user MIMO Systems}
\author{{Keke Zu, Bin Song, Martin Haardt, {\it Senior Member, IEEE} and Rodrigo C. de Lamare, {\it Senior Member, IEEE}} }

\maketitle\thispagestyle{empty} \vspace*{-1.5em}

\begin{abstract}
Tomlinson-Harashima precoding (THP) is a nonlinear processing technique employed at the transmit side to implement the concept of dirty paper coding (DPC). The perform of THP, however, is restricted by the dimensionality constraint that the number of transmit antennas has to be greater or equal to the total number of receive antennas.
In this paper, we propose an iterative coordinate THP algorithm for the scenarios in which the total number of receive antennas is larger than the number of transmit antennas. The proposed algorithm is implemented on two types of THP structures, the decentralized THP (dTHP) with diagonal weighted filters at the receivers of the users, and the centralized THP (cTHP) with diagonal weighted filter at the transmitter.
Simulation results show that a much better bit error rate (BER) and sum-rate performances can be achieved by the proposed iterative coordinate THP compared to the previous linear art.
\end{abstract}


\section{Introduction}
For the Multi-user MIMO (MU-MIMO) broadcast channel with independent and identically distributed (i.i.d.) Gaussian
interference, its capacity region is achieved by the dirty paper coding (DPC) \cite{Spencer, Shamai}. The concept of DPC is originally proposed in \cite{Costa} that the capacity of the broadcast channel is the same as if the interference was not present by setting the transmitted signal equal to the desired data minus the interference. The DPC theory, however, is not suitable for
practical application due to the requirement of infinitely long codewords \cite{Khina}.

Tomlinson-Harashima precoding (THP) \cite{Tomlinson, Harashima} is a pre-equalization technique originally proposed for
channels with intersymbol interference (ISI). Then, the THP technique was extended from temporal equalization to spatial equalization for MIMO precoding in \cite{Fischer,Christoph}, where an equal number of transmit and receive antennas is assumed. Although THP still suffers a performance loss as refer to the theoretical upper bound achieved by DPC \cite{WY}, it can work as a cost-effective replacement of DPC in practice by implementing a LQ decomposition \cite{Erez, Keke}. In the literature, however, the system dimensionality for implementing the THP algorithms is always set as the number of transmit antennas greater than or equal to the total number of receive antennas, which is termed as the dimensionality constraint in \cite{Spencer02}.

To overcome the dimensionality constraint for the precoding algorithms, a receive antenna selection method is proposed in \cite{Shen}.  Another approach is the coordinated beamforming (CBF) which employs iterative operations to jointly update the transmit-receive beamforming vectors \cite{Zhou}-\cite{Bin03}. The convergence behavior of the iterations is not considered in \cite{Zhou}, and the coordinated transmission strategy in \cite{Chae} only supports a single data stream to each user. A flexible coordinated beamforming (FlexCoBF) algorithm is proposed in \cite{Bin}-\cite{Bin03} to support the transmission of multiple data streams to each user for linear precoding techniques \cite{Michael}.

Based on our previous work on FlexCoBF, an iterative coordinate nonlinear THP algorithm is proposed in this work.
There are two THP structures according to the positions of the diagonal weighted filter, decentralized filters located at the receivers or centralized filters deployed at the transmitter, which are denoted as dTHP or cTHP, respectively \cite{MH, Keke02}. Most of the previous research works on THP, however, have only focused on one of the structures.
In this work, we develop the iterative coordinate THP for both of the two THP structures. The main contributions of the work can be summarized as
\begin{enumerate}
   \item An iterative coordinate strategy is developed to solve the dimensionality constraint for the nonlinear THP algorithms.
     \item The proposed iterative coordinate THP algorithm supports the transmission of multiple data streams.
     \item The proposed iterative coordinate THP algorithm is developed for both cTHP and dTHP structures.
   \item Not only the sum-rate but also the BER performance are investigated.
 \end{enumerate}

This paper is organized as follows. The system model and the basics of THP techniques are described in Section II. The proposed iterative coordinated THP algorithm is described in detail in Section III. Simulation results and conclusions are presented in Section IV and Section V, respectively.

\section{System Model and THP Algorithms}
\subsection{The MU-MIMO System Model}
\begin{figure}[h]
\begin{center}
\def\epsfsize#1#2{0.85\columnwidth}
\epsfbox{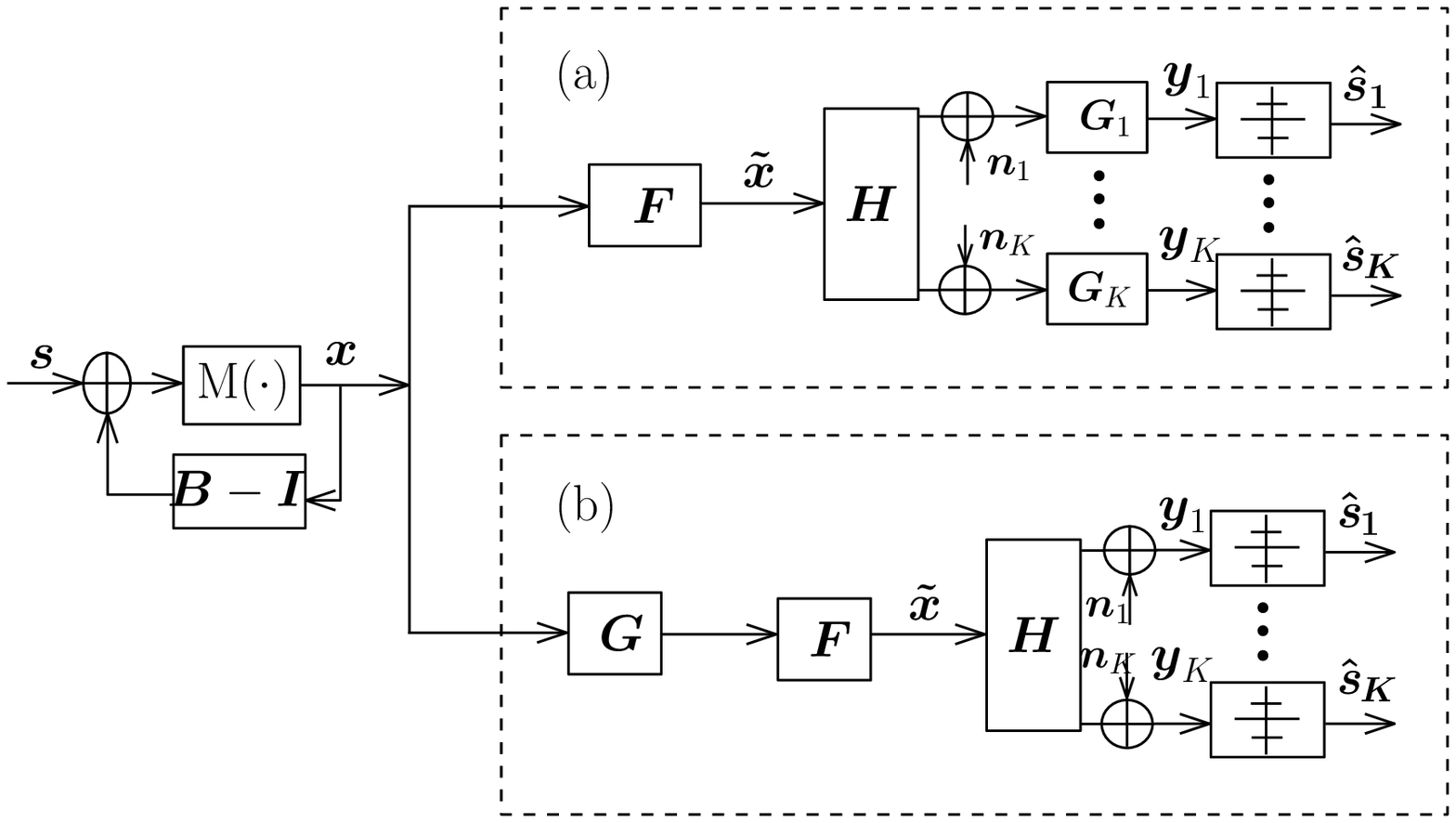} % \vspace{-0.5em}
\caption{\footnotesize The MU-MIMO System Model\\} \label{MU_MIMO_System_Model}
\end{center}
\end{figure}

We consider an uncoded MU-MIMO broadcast system illustrated in Fig. \ref{MU_MIMO_System_Model} , with $N_t$ transmit antennas
equipped at the base station (BS), $K$ users in the system each equipped with $N_k$ receive antennas, and the total number
of receive antennas is $N_r=\sum _{k=1}^{K}N_k$. The combined transmit data streams are denoted as ${\boldsymbol s}=[{\boldsymbol s^T_1},{\boldsymbol s^T_2},\cdots,{\boldsymbol s^T_K}]^T\in\mathbb{C}^{r\times 1}$ with $\boldsymbol s_k\in\mathbb{C}^{r_k\times 1}$, where $r$ is the total number of transmit data streams and $r_k$ is the number of $k$th user's transmit data streams.  The combined channel matrix is denoted as ${\boldsymbol H}=[{\boldsymbol H^T_1},{\boldsymbol H^T_2},\cdots,{\boldsymbol H^T_K}]^T\in\mathbb{C}^{N_r\times N_t}$ and $\boldsymbol H_k\in\mathbb{C}^{N_k\times N_t}$ is the $k$th user's channel matrix.

Note that power-loading schemes \cite{Christoph} could be used to
determine the number of data streams or allocate more power to a
weaker user to improve the overall performance. However, for
simplicity, we assume that no power loading between users and streams is performed.

\subsection{Two Basic THP Structures}
Based on the knowledge of CSI at the transmit side, the interference of the parallel streams of a MU-MIMO system with spatial
multiplexing can be subtracted from the current stream. This successive inter-user interference cancellation technique at the transmit side is known as THP. Generally, there are three filters to implement THP algorithm: the feedback filter $\boldsymbol B$, the feedforward filter $\boldsymbol F$, and the scaling filter $\boldsymbol G$.
According to the position of $\boldsymbol G$, there are two THP structures, which are illustrated in Fig. \ref {Two_THP_Structures} below. The decentralized THP (dTHP) employs $\boldsymbol G$ (or sub-matrices of it) at the receivers, whereas the centralized THP (cTHP) uses $\boldsymbol G$ at the transmitter.

\begin{figure}[htp]
\begin{center}
\def\epsfsize#1#2{0.95\columnwidth}
\epsfbox{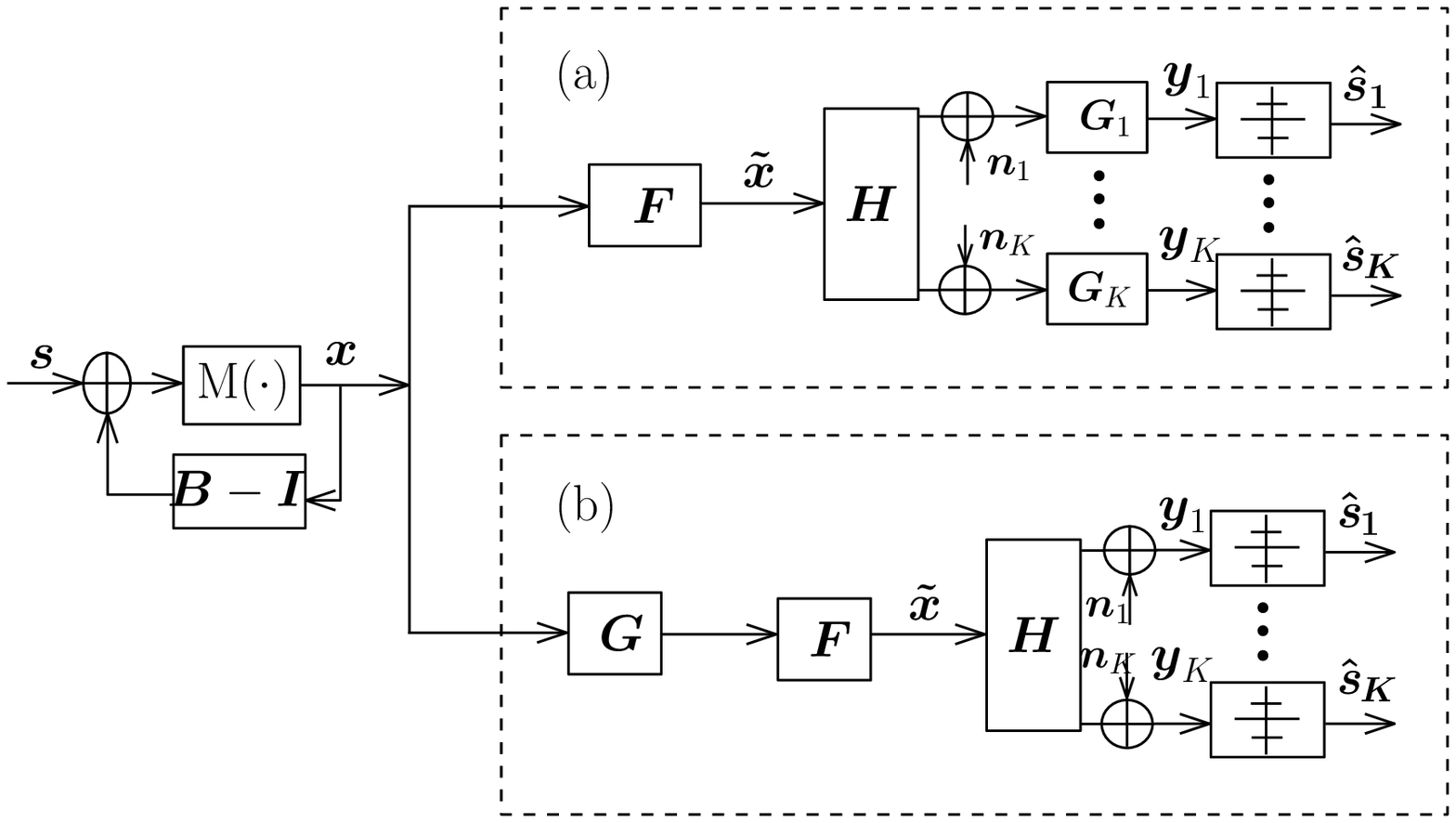} %\vspace{-0.8em}
\caption{\footnotesize The two basic THP structures\\
(a) Decentralized THP: the scaling matrix $\boldsymbol G$ is separately placed at the receivers.\\
(b) Centralized THP: the scaling matrix $\boldsymbol G$ is placed at the transmitter.} \label{Two_THP_Structures}
\end{center}
\end{figure}

The feedback filter $\boldsymbol B$ is used to successively cancel the interference caused by the previous streams from the current
stream. Therefore, the feedback filter $\boldsymbol B$ should be a lower triangular matrix with ones on the main diagonal \cite{Christoph}.
The feedforward filter $\boldsymbol F$ is used to enforce the spatial causality and has to be implemented at the transmit side for MU-MIMO systems because the physically distributed users cannot be processed jointly. The scaling filter $\boldsymbol G$ contains the corresponding weighted coefficient for each stream and thus it should have a diagonal structure. The quantity $\boldsymbol x$ is the combined transmit signal vector after the feedback operation and $\tilde {\boldsymbol x}$ is the combined transmit signal vector after precoding,
\begin{align}
{\tilde {\boldsymbol x}}^{(dTHP)}=\boldsymbol F{\boldsymbol x} ~{\rm and}~ {\tilde {\boldsymbol x}}^{(cTHP)}=\boldsymbol F\boldsymbol G{\boldsymbol x}.
\end{align}
The filters of THP can be effectively obtained by implementing an LQ decomposition \cite{Keke02} on the combined channel matrix $\boldsymbol H$, that is
\begin{eqnarray}
\boldsymbol {H}=\boldsymbol L\boldsymbol Q,
\end{eqnarray}
where $\boldsymbol L$ is a lower triangular matrix and $\boldsymbol Q$ is a unitary matrix (by unitary we mean $\boldsymbol Q^{H}\boldsymbol Q=\boldsymbol Q\boldsymbol Q^{H}=\boldsymbol I$). Therefore, the filters for the THP algorithms can be obtained as
\begin{align}
\boldsymbol F={\boldsymbol Q}^H,\\
\boldsymbol G={\rm {diag}}[l_{1,1},l_{2,2},\cdots,l_{S,S}]^{-1},\\
\boldsymbol B^{\rm (dTHP)}=\boldsymbol G\boldsymbol L, \boldsymbol B^{\rm (cTHP)}=\boldsymbol L\boldsymbol G,
\end{align}
where $l_{i,i}$ is the $i$th diagonal element of the matrix $\boldsymbol L$.

From Fig. \ref {Two_THP_Structures}, the transmitted symbols $x_i$ are successively generated as
\begin{align}
 x_i={\rm M}({s_i- \sum _{j=1}^{i-1}b_{i,j}{x_j}}),~ {i=1,\cdots,r},
\end{align}
where $s_i$ is the $i$th transmit data stream with variance $\sigma_s^2$ and $b_{i,j}$ are the elements of $\boldsymbol B$ in row $i$ and column $j$. The transmit power is significantly increased as the amplitude of $x_i$ exceeds the modulation boundary by implementing the successive cancellation. A modulo operation ${\rm M}(\cdot)$, which is defined element-wise, is employed to reduce the amplitude of the channel symbol $x_i$ to the boundary of the modulation alphabet \cite{Dietrich}
\begin{align}
{\rm M}(x_i)=x_i- \biggl\lfloor {{\rm{Re}}(x_i)\over \tau }+{1\over 2}\biggr\rfloor\tau-j\biggl\lfloor {{\rm{Im}}(x_i)\over \tau }+{1\over 2}\biggr\rfloor\tau,
\end{align}
where $\tau$ is a constant for the periodic extension of the constellation. The specific value of $\tau$ depends on the chosen modulation alphabet. Common choices for $\tau$ are $\tau=2\sqrt 2$ for QPSK symbols and $\tau=8\sqrt {10}$ in case of rectangular 16-QAM when the symbol variance is one. The modulo processing is equivalent to adding a perturbation vector $\boldsymbol d$ to the transmit data $\boldsymbol s$, such that the modified transmit data are \cite{Keke02}
\begin{align}
\boldsymbol v=\boldsymbol s +\boldsymbol d.
\end{align}
Thus, the initial signal constellation is extended periodically and the effective $k$th transmit data symbols $\boldsymbol v_k$ are taken from the expanded set.

The received signal for dTHP and cTHP, after the feedback, feedforward and the scaling filters, is respectively given by
\begin{eqnarray}
\label{Rx_cTHP}{\boldsymbol y}^{\rm (cTHP)}=\beta({\boldsymbol H}\cdot{1\over\beta}\boldsymbol F\boldsymbol G{\boldsymbol x} +\boldsymbol n),
\end{eqnarray}

\begin{eqnarray}
\label{Rx_dTHP}{\boldsymbol y}^{\rm (dTHP)}=\boldsymbol G({\boldsymbol H}\boldsymbol F{\boldsymbol x} +\boldsymbol n),
\end{eqnarray}
where the quantity $\boldsymbol n=[{\boldsymbol n^T_1},{\boldsymbol n^T_2},\cdots,{\boldsymbol n^T_K}]^T$ is the combined Gaussian noise vector with i.i.d. entries of zero mean and variance $\sigma_n^2$.
The factor $\beta$ is used to impose the power constraint ${\rm E}\| \tilde {\boldsymbol x}\|^2=\xi$ with $\xi$ being the average transmit power.

\section{Proposed Iterative Coordinated THP Algorithm}
The above discussed original THP algorithms are only feasible under the dimensionality constraint that $N_t\geq N_r$. In this section, the coordinate concept is developed for THP algorithms to overcome the dimensionality constraint.

For the case when $N_r > N_t$, the MU-MIMO system cannot support the transmission of $N_r$ data streams. Assume the number of actually transmitted data streams is $r$ and it should satisfy $r\leq N_t$, which the physical meaning is that the number of maximum transmitted data streams cannot beyond the number of transmit antennas $N_t$. Then, a coordinate filter $\boldsymbol W_k\in\mathbb{C}^{r\times N_k}$ is introduced at each user to adjust the transmit-receive filters iteratively. The equivalent channel matrix ${\boldsymbol H_e}\in\mathbb{C}^{r\times N_t}$ is obtained as
\begin{align}
\boldsymbol  H_e=\begin{bmatrix} \boldsymbol W_1 & 0 & \ldots & 0 \\ 0 & \boldsymbol W_2& \ldots & 0 \\ \vdots & \vdots & \ddots &
\vdots\\ 0 & 0 & 0 & \boldsymbol W_K \end{bmatrix} \begin{bmatrix} \boldsymbol H_1 \\ \boldsymbol H_2\\ \vdots  \\ \boldsymbol H_K \end {bmatrix}=\begin {bmatrix} \boldsymbol W_1\boldsymbol H_1 \\ \boldsymbol W_2\boldsymbol H_2\\ \vdots  \\ \boldsymbol W_K\boldsymbol H_K \end {bmatrix}.
\end{align}

In Fig. \ref{Two_THP_Structures}, the feedback processing is mathematically equivalent to an inversion operation ${\boldsymbol B}^{-1}$. Therefore, the transmitted symbol $\boldsymbol x$ can be rewritten as
\begin{align}
\label{Equivalent_V}\boldsymbol x={\boldsymbol B}^{-1}\boldsymbol v={\boldsymbol B}^{-1}(\boldsymbol s +\boldsymbol d),
\end{align}
Substitute the equation (\ref{Equivalent_V}) into equations (\ref{Rx_cTHP}) and (\ref{Rx_dTHP}), the received signal for cTHP and dTHP can be respectively rewritten as
\begin{eqnarray}
 \label{Parallel_cTHP}{{\boldsymbol r}^{\rm (cTHP)}}&=&\boldsymbol v +\beta\boldsymbol n,
\end{eqnarray}
\begin{eqnarray}
 \label{Parallel_dTHP}{{\boldsymbol r}^{\rm (dTHP)}}&=&\boldsymbol v +\boldsymbol G\boldsymbol n.
\end{eqnarray}
From the above equations (\ref{Parallel_cTHP}) and (\ref{Parallel_dTHP}), the MU-MIMO broadcast channel is transformed into effective parallel single-user MIMO (SU-MIMO) channels. Therefore, the multi-user interference (MUI) is enforced to zero at each user by the successively THP processing. In the proposed iterative coordinate THP algorithm, the coordinate filter $\boldsymbol W_k$ at each receiver is initialized with random matrices. Then, iterative computations are employed to update the coordinate filter $\boldsymbol W_k$ to enforce zero MUI constraint as suggested in equations (\ref{Parallel_cTHP}) and (\ref{Parallel_dTHP}). Assume the variable $p$ represents the iteration index, the proposed iterative coordinate THP algorithm is performed in 5 steps.

\begin{enumerate}
   \item Initialize the iteration index $p$ to zero and $\boldsymbol W_k^{(0)}$ to random matrices. Set the constant $\epsilon$ as the threshold for iteratively enforce zero MUI constraint.
     \item Set $p=p+1$ and compute the equivalent channel matrix $\boldsymbol H_e^{(p)}$ as
          \begin{align}
                    \boldsymbol H_e^{(p)}=\begin {bmatrix} \boldsymbol W_1^{(p-1)}\boldsymbol H_1\\ \boldsymbol W_2^{(p-1)}\boldsymbol H_2\\ \vdots  \\ \boldsymbol W_K^{(p-1)}\boldsymbol H_K\end {bmatrix}.\nonumber
                \end{align}
     \item Apply the LQ decomposition on the equivalent channel matrix $\boldsymbol H_e^{(p)}=\boldsymbol L_e^{(p)}\boldsymbol Q_e^{(p)}$ to obtain the THP filters as
           \begin{align}
          \boldsymbol  F_e^{(p)}={\boldsymbol Q_e^{(p)}}^H,\nonumber\\
          \boldsymbol  G_e^{(p)}={\rm {diag}}(\boldsymbol L_e^{(p)})^{-1},\nonumber\\
                    \boldsymbol {B_e^{(p)}}^{\rm (cTHP)}=\boldsymbol L_e^{(p)}\boldsymbol G_e^{(p)},\nonumber\\
          \boldsymbol {B_e^{(p)}}^{\rm (dTHP)}=\boldsymbol G_e^{(p)}\boldsymbol L_e^{(p)}. \nonumber
         \end{align}
   \item Update the $p$th combined coordinate filter $\boldsymbol W^{(p)}$ as
             \begin{align}
                    \boldsymbol {W_e^{(p)}}^{\rm (cTHP)}=\boldsymbol H\boldsymbol  F_e^{(p)}\boldsymbol G_e^{(p)}{\boldsymbol {B_e^{(p)}}^{\rm (cTHP)}}^{-1},\nonumber\\
          \boldsymbol {W_e^{(p)}}^{\rm (dTHP)}=\boldsymbol H\boldsymbol  F_e^{(p)}{\boldsymbol {B_e^{(p)}}^{\rm (dTHP)}}^{-1}. \nonumber
         \end{align}
    \item Track the alterations of the residual MUI after the THP processing as
               \begin{align}
                    {\rm MUI} (\boldsymbol H_e^{(p+1)}\boldsymbol {P_e^{(p)}}^{\rm (cTHP)})=\|{\rm off}(\boldsymbol H_e^{(p+1)}\boldsymbol {P_e^{(p)}}^{\rm (cTHP)})\|, \nonumber\\
                    {\rm MUI} (\boldsymbol H_e^{(p+1)}\boldsymbol {P_e^{(p)}}^{\rm (dTHP)})=\|{\rm off}(\boldsymbol H_e^{(p+1)}\boldsymbol {P_e^{(p)}}^{\rm (dTHP)})\|, \nonumber
         \end{align}
                where ${P_e^{(p)}}^{\rm (cTHP)}=\boldsymbol F_e^{(p)}\boldsymbol G_e^{(p)}{\boldsymbol {B_e^{(p)}}^{\rm (cTHP)}}^{-1}$, ${P_e^{(p)}}^{\rm (dTHP)}=\boldsymbol F_e^{(p)}{\boldsymbol {B_e^{(p)}}^{\rm (dTHP)}}^{-1}$ and ${\rm off}(\boldsymbol A)$ denotes the off-diagonal elements of the matrix $\boldsymbol A$.
                If the residual MUI is above the threshold $\epsilon$, go back to step 2. Otherwise, convergence is achieved and the iterative procedure can be ended.
 \end{enumerate}

For setting the value of the threshold $\epsilon$, we usually apply $\epsilon=10^{-5}$ to effectively perform the proposed algorithm and also it is not necessary to set the threshold $\epsilon$ exactly equals to zero.
% The pseudo-code of the proposed iterative cooperate THP algorithm is summarized in Table I.

\section{Simulation Results}
In this section, we assess the performance of the proposed iterative coordinate THP algorithms. A system with $N_t=8$ transmit antennas and $K=4$ users each equipped with $N_k=3$ receive antennas is considered;
this scenario is denoted as the $(3,3,3,3)\times 8$ case. The quantity $E_b/N_0$ is defined as $E_b/N_0={N_r E_s\over N_tN\sigma_n^2}$ with $N$ being the number of information bits transmitted per channel
symbol. Uncoded QPSK modulation scheme is employed in the simulations. The threshold $\epsilon$ is set to $10^{-5}$, and the maximum iteration number is restricted to $50$. The channel matrix ${\boldsymbol H}$ is assumed to be a complex i.i.d. Gaussian matrix with zero mean and unit variance.

\begin{figure}[htp]
\begin{center}
\def\epsfsize#1#2{0.95\columnwidth}
\epsfbox{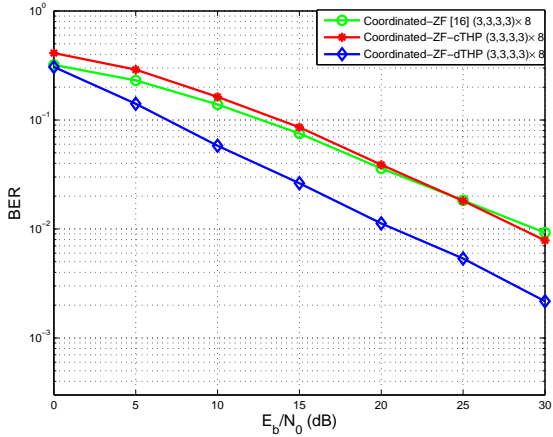} % \vspace{-0.5em}
\caption{\footnotesize The Comparison of BER performance with QPSK\\} \label{Co_THP_BER}
\end{center}
\end{figure}

For the $(3,3,3,3)\times 8$ MU-MIMO broadcast channel, the proposed iterative coordinate THP is implemented to support the maximum data streams at each user $r_k=2$. The normal case of $(2,2,2,2)\times 8$ MU-MIMO broadcast channel is also simulated for comparison. The BER performance is illustrated in Fig. \ref{Co_THP_BER}, from which we can find out that the precoding algorithms with iterative coordinate process suffer a performance loss compared to their counterparts implemented with normal case. The reason is that the introduced coordinate filter $\boldsymbol W$ at each user brought extra noise enhancement for relaxing the dimensionality constraint. The proposed iterative coordinate dTHP achieves a gain of more than $8$ dB at BER of $10^{-2}$ as compared to our previously proposed linear iterative coordinate ZF precoding. The iterative coordinate cTHP can still work well in this overloaded system but its performance is not so good as that of linear one at low $E_b/N_0$s.

\begin{figure}[htp]
\begin{center}
\def\epsfsize#1#2{0.95\columnwidth}
\epsfbox{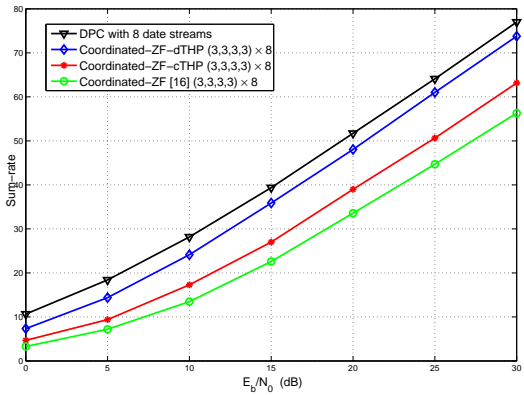} % \vspace{-0.5em}
\caption{\footnotesize The Comparison of Sum-rate performance\\} \label{Co_THP_Sumrate}
\end{center}
\end{figure}

The sum-rate performance is displayed in Fig. \ref{Co_THP_Sumrate}. A much better sum-rate performance is achieved by both of the proposed iterative coordinate cTHP and dTHP than our previously proposed linear ZF precoding, and the sum-rate performance of iterative coordinate dTHP is very close to that of the DPC. Compared to the original dTHP implemented in the normal case, only a small loss when $E_b/N_0$ is over $20$ dB is performed. For the iterative coordinate cTHP, its sum-rate performance is not so sensitive as its BER performance compared to the original cTHP. Therefore, we have found that better BER and sum-rate performances can be achieved by the iterative coordinate dTHP as compared to the iterative coordinate cTHP. The BER performance of the iterative coordinate cTHP is more sensitive to the coordinate filter $\boldsymbol W$ but its sum-rate performance maintains the robustness.

\section{conclusion}

In this paper, an iterative coordinate THP algorithm have been proposed to relax the dimensionality constraint suffered by the original THP algorithms. Therefore, we consider the scenarios where $N_r > N_t$. This condition is fulfilled in many scenarios that have been studied recently. For example, the users across cell borders have to be considered jointly by base stations (BSs) for coordinated multi-point (CoMP) transmission. Furthermore, when each user is equipped with multiple antennas, the BS simultaneously serves as many users as possible, which corresponds to a large number of receive antennas.

\end{document}